\documentclass[utf8]{FrontiersinHarvard}

\usepackage{graphicx}
\usepackage{amsmath}
\usepackage{amsfonts}
\usepackage{amssymb}
\usepackage{times,txfonts}
\usepackage{braket}
\usepackage[colorlinks=true,linkcolor=blue,urlcolor=blue,citecolor=blue]{hyperref}
\usepackage{xcolor}
\usepackage[onehalfspacing]{setspace}

\def\keyFont{\fontsize{8}{11}\helveticabold }
\def\firstAuthorLast{Samuel L. Jacob {et~al.}} %use et al only if is more than 1 author
\def\Authors{Samuel L. Jacob\,$^{1,*}$, Laetitia P. Bettmann\,$^{1}$, Artur M. Lacerda\,$^{1}$, Krissia Zawadzki\,$^{2}$, Stephen R. Clark\,$^{3}$, John Goold\,$^{1,4}$, Juan José Mendoza-Arenas\,$^{3,5,6}$}
\def\Address{$^{1}$School of Physics, Trinity College Dublin, College Green, Dublin 2, D02 K8N4, Ireland \\
$^{2}$Instituto de Física de São Carlos, Universidade de São Paulo,
CP 369, 13560-970 São Carlos, São Paulo, Brazil \\
$^{3}$H. H. Wills Physics Laboratory, University of Bristol, Bristol BS8 1TL, United Kingdom \\
$^{4}$Trinity Quantum Alliance, Unit 16, Trinity Technology and Enterprise Centre, Pearse Street, Dublin 2, D02 YN67, Ireland \\
$^{5}$Department of Mechanical Engineering and Materials Science, University of Pittsburgh, Pittsburgh,
Pennsylvania 15261, USA \\
$^{6}$Department of Physics and Astronomy, University of Pittsburgh, Pittsburgh, Pennsylvania 15261, USA}

\def\corrAuthor{Samuel L. Jacob}

\def\corrEmail{samjac91@gmail.com}

\begin{document}
\onecolumn
\firstpage{1}

\title[Dephasing-assisted transport with a linear potential]{Dephasing-assisted transport in a tight-binding chain with a linear potential} 

\author[\firstAuthorLast]{\Authors} %This field will be automatically populated
\address{} %This field will be automatically populated
\correspondance{} %This field will be automatically populated

\extraAuth{}% If there are more than 1 corresponding author, comment this line and uncomment the next one.
%\extraAuth{Laetitia P. Bettmann \\ Laboratory X2, Institute X2, Department X2, Organization X2, Street X2, City X2 , State XX2 (only USA, Canada and Australia), Zip Code2, X2 Country X2, email2@uni2.edu}

\maketitle

\begin{abstract}
\section{}

An environment interacting with a quantum system can enhance transport through the suppression of quantum effects responsible for localization. In this paper, we study the interplay between bulk dephasing and a linear potential in a boundary-driven tight-binding chain. A linear potential induces Wannier-Stark localization in the absence of noise, while dephasing induces diffusive transport in the absence of a tilt. We derive an approximate expression for the steady-state current as a function of both dephasing and tilt which closely matches the exact solution for a wide range of parameters. From it, we find that the maximum current occurs for a dephasing rate equal to the period of Bloch oscillations in the Wannier-Stark localized system. We also find that the current displays a maximum as a function of the system size, provided that the total potential tilt across the chain remains constant. Our results can be verified in current experimental platforms and represents a step forward in analytical studies of environment-assisted transport.

\tiny
 \keyFont{ \section{Keywords:} Dephasing-assisted transport, Wannier-Stark localization}

\end{abstract}

\section{Introduction}

The quantum features of an open system interacting with a macroscopic environment are inevitably destroyed in a process known as decoherence \cite{Joos2003,Breuer2007,Rivas2012}. However, the notion that the environment is always detrimental for quantum processes -- such as information processing and transport -- has been challenged for more than a decade, prompted by investigations of quantum effects in biological systems \cite{Plenio2008,Mohseni2008,Rebentrost2009,Chin2010}. It is now understood that the environment can assist energy transport in non-interacting quantum systems \cite{Plenio2008,Mohseni2008,Rebentrost2009,Chin2010,Sinayskiy2012,Harush2020,Harush2021,Alterman2024,Ferreira2024}, an effect which has been experimentally verified in quantum networks of photons \cite{Viciani2015,Tang2024}, trapped ions \cite{Gorman2018,Maier2019} and superconducting circuits \cite{Potocnik2018}. Although several mechanisms for environment-assisted quantum transport have been proposed and debated \cite{Rebentrost2009,Chin2010,Harush2020}, a clear mechanism is at play in localized quantum systems -- when destructive interference responsible for localization and transport suppression is destroyed by the environment, quantum transport is enhanced. This is indeed the expected impact of environmental coupling on Anderson and Wannier-Stark localization. In Anderson localized systems, quantum transport is suppressed as a consequence of lattice disorder \cite{Anderson1958}. On the other hand, Wannier-Stark localization occurs in the presence of a linear lattice potential such as an electric field \cite{Bloch1929,Zener1934,Wannier1962}; in this case, coherent (Bloch) oscillations take place within the region of localization \cite{Bloch1929,Hartmann2004,Nieuwenburg2019,Guo2021}. Despite the fact that Wannier-Stark localized systems subject to noise have been investigated \cite{Burkhardt2013,Bhakuni2019,Langlett2023,Teretenkov2024}, the literature on this subject is still scarce.

Environment-assisted transport can also affect the non-equilibrium transport properties of many-body quantum systems. A common approach to evidence this effect is to drive the quantum system to a non-equilibrium steady state by high temperature reservoirs located at the boundaries \cite{Bertini2021,Landi2022}. The dephasing effect of the environment renders the transport diffusive, which has been shown to enhance the steady-state current in boundary-driven systems with disorder \cite{Znidaric2010a,Znidaric2013,Znidaric2016} or quasi-periodicity \cite{Lacerda2021} and to modify the transport properties of systems with long-range hopping \cite{Sarkar2024}. Moreover, Markovian boundary-driven non-interacting systems are amenable to analytical treatment even in the presence of dephasing, with exact and approximate solutions of steady-state quantities available for tight-binding chains \cite{Znidaric2010,Znidaric2010a,Znidaric2011,Znidaric2013,Turkeshi2021}.

In this paper, we study dephasing-assisted transport in a boundary-driven, tight-binding chain with a linear potential. We make use of a steady-state ansatz elaborated in Refs.~\cite{Znidaric2010,Znidaric2010a,Znidaric2011,Znidaric2013} to numerically access the relevant steady-state quantities for up to one thousand chain sites. By performing a minimal approximation, we derive an analytical expression for the steady-state current which matches the exact dynamics for a vast range of parameters. From it, we find that the current is maximized at a dephasing rate equal to the period of Bloch oscillations in a Wannier-Stark localized system. We also find that the current displays a maximum as a function of the system size, provided that the total potential tilt across the chain is kept constant. Our results present a significant contribution in the analytics of environment-assisted transport and can be experimentally verified in several platforms.

The paper is organized as follows. In Sec.~\ref{sec:setup} we introduce the model and the steady-state ansatz used in our study. In Sec.~\ref{sec:dat} we present and discuss our results for dephasing-assisted transport. In Sec.~\ref{sec:conclusion} we present the conclusion and outlook of our work. Sec.~\ref{app:suppmat} is the supplementary material, containing technical details referenced throughout the main text. 

\section{Setup}
\label{sec:setup}

\subsection{The model}
\label{sec:model}
We study a non-interacting, one-dimensional lattice of $L$ sites with Hamiltonian given by the $XX$ model
\begin{align}
    \label{hamiltonian}
    H = J \sum_{j=1}^{L-1}~ (\sigma_{j}^x \sigma^{x}_{j+1} + \sigma_{j}^y \sigma^{y}_{j+1}) + \sum_{j=1}^{L} \epsilon_j \sigma_j^z \; .
\end{align}
Here, $\{ \sigma^{x}_j, \sigma^{y}_j, \sigma^{z}_j \}_{j=1}^{L}$ are Pauli matrices, $J$ is the hopping term and $\epsilon_j$ is a local field which will be specified shortly. In order to induce a non-equilibrium state in the system, we couple the boundary sites to high-temperature reservoirs with
different chemical potentials; in addition, each site is also exposed to its own local reservoir which induces dephasing. All the reservoirs are assumed to be ideal and couple weakly to the system. The dynamics of the system's density operator $\rho(t)$ at time $t$ is given by the Lindblad master equation \cite{Breuer2007,Rivas2012,Landi2022}
\begin{align}
    \label{masterequation}
    \frac{d \rho(t)}{dt} = -\frac{i}{\hbar}[H,\rho(t)] + \sum_{\alpha = l,r,d}\mathcal{L}^{\mathrm{\alpha}}[\rho(t)] \; .
\end{align}
Here the superoperator $\mathcal{L}^{\alpha}$ describes dissipation induced by the reservoirs $\alpha = l,r,d$ corresponding to left boundary, right boundary and dephasing, respectively. It is a sum of local Lindblad jump operators of the form
\begin{align}
    \mathcal{L}^{\alpha}~ [\cdot] = \sum_{j,\beta} L^{\alpha \beta}_j \cdot L_j^{\alpha \beta \dagger} - \frac{1}{2} \big\{ L_j^{\alpha \beta \dagger} L^{\alpha \beta}_j, \cdot \big\} \; ,
\end{align}
where $j$ labels the site acted on by the reservoir $\alpha$ and $\beta$ labels different jump operators on that same site. The effect of each boundary reservoir ($\alpha = l,r$) is represented by two jump operators
\begin{align}
    \label{boundaryoperators}
    L^{l \pm}_1 = \sqrt{\frac{\Gamma(1 \pm f)}{2}} \sigma_1^{\pm} \hspace{2.5mm} , \hspace{2.5mm} L^{r\pm}_L = \sqrt{\frac{\Gamma(1 \mp f)}{2}} \sigma_L^{\pm} \; ,
\end{align}
where $\Gamma$ is the coupling rate at the boundaries, $f$ is the chemical potential bias and $\sigma_j^{\pm} = (\sigma_j^x \pm i \sigma_j^y)/2$. For forward bias ($0 \leq f \leq 1$) excitations are mostly created on the first site and annihilated on the last site; the opposite reasoning applies for reverse bias ($-1 \leq f \leq 0$). In order to describe the effects of dephasing, we consider $L$ jump operators acting on individual sites as if each of them were coupled to its own reservoir
\begin{align}
    L_j^{d} = \sqrt{\frac{\gamma}{2}} \sigma^{z}_j \hspace{2.5mm} \; , \hspace{2.5mm} j = 1, 2, ..., L \; ,
\end{align}
where $\gamma$ is the dephasing rate. 

The key observables in the system are the magnetization and its associated current. We denote by $\langle A \rangle(t) \equiv \mathrm{Tr}[A \rho(t)]$ the expectation value of an observable $A$ at time $t$. Differentiating with respect to time and using Eq.~\eqref{masterequation} yields a continuity equation (see Sec.~\ref{app:heisenberg} for more details)
\begin{align}
    \label{heisenbergequation}
    \frac{d \langle \sigma^z_j \rangle(t)}{dt} & = \langle I_{j-1} \rangle(t) - \langle I_{j} \rangle(t) + \delta_{j,1} \Gamma [f - \langle \sigma^z_j \rangle(t)] - \delta_{j,L} \Gamma [f + \langle \sigma^z_j \rangle(t)] \; \; ,
\end{align}
involving  the expectation value of the magnetization current $\langle I_j \rangle(t)$  flowing from $j$ to $j+1$, where
\begin{align}
    \label{currentoperator}
    I_{j} = \frac{2J}{\hbar} (\sigma_{j}^{x} \sigma_{j+1}^{y} - \sigma_{j}^{y} \sigma_{j+1}^{x}) \;
\end{align}
is the corresponding current operator. Note that the dephasing reservoir does not change the average magnetization and therefore does not contribute explicitly to Eq.~\eqref{heisenbergequation}, although it still affects the average magnetization and currents implicitly. Moreover, the boundary reservoirs do not contribute to the expression of the current in the bulk of the chain. 

\begin{figure}[t]
  \centering
  \includegraphics[width=0.7\textwidth]{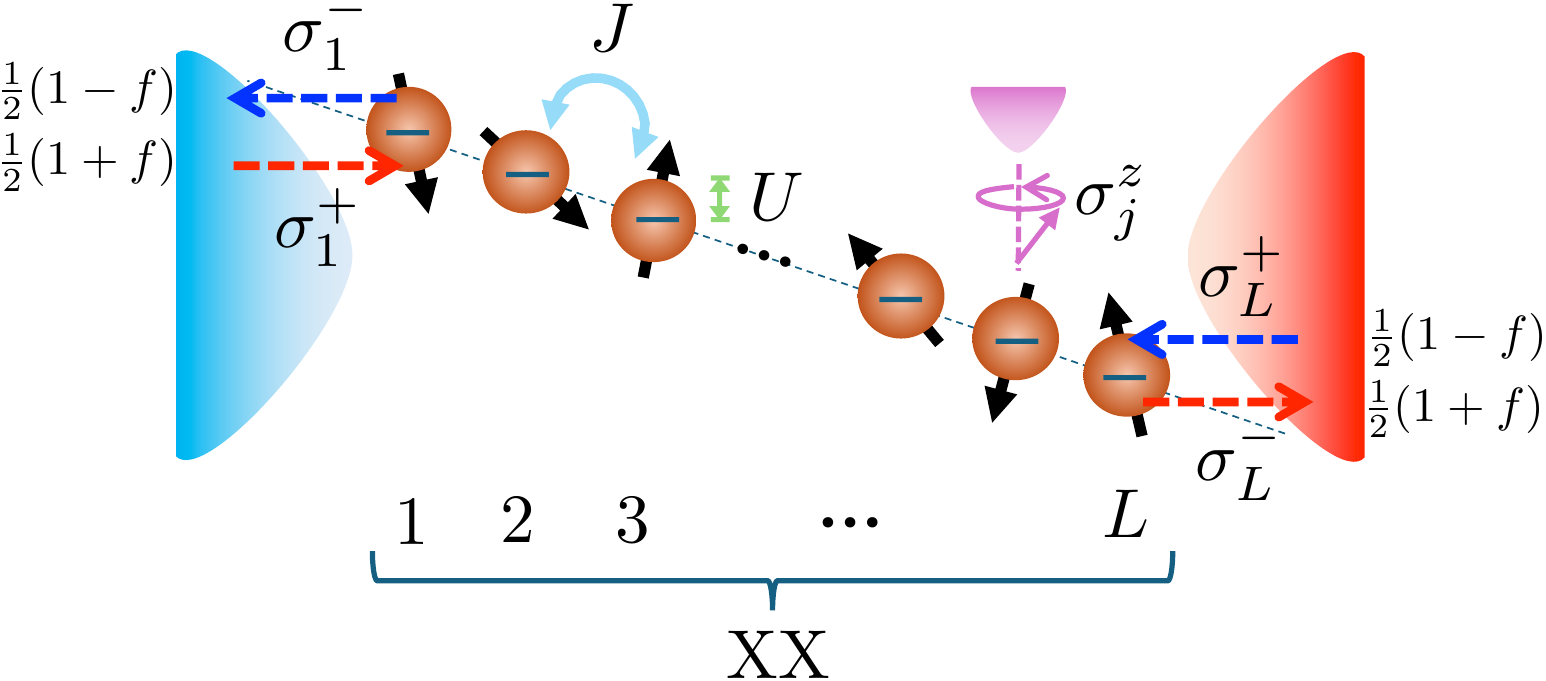}
  \caption{A schematic diagram of the system under study. The lattice is modeled as an $XX$ spin chain, characterised by a hopping amplitude $J$ and on-site field linearly decaying in steps of $U$ across the chain. The boundary sites are coupled to separate reservoirs that inject/eject spin excitations controlled by a bias $f$, while in the bulk each spin is coupled to its own local reservoir which induces dephasing.}
  \label{fig:schematic}
\end{figure}

When the steady state is reached, the left-hand side of Eqs.~\eqref{masterequation} and \eqref{heisenbergequation} vanishes. We denote by $\langle A \rangle_{\infty} \equiv \mathrm{Tr}[A \rho(\infty)]$ the expectation value of $A$ at the steady state $\rho(\infty)$. From Eq.~\eqref{heisenbergequation} we obtain the following equations for the steady-state current
\begin{align}
    \label{currentboundaryleft}
    \langle I_1 \rangle_{\infty} & = \Gamma (f - \langle \sigma^z_1 \rangle_{\infty}) \; , \\
    \langle I_{j-1} \rangle_{\infty} & = \langle I_{j} \rangle_{\infty} \; , \hspace{15mm} j = 2, 3, \dots, L-1 \\
    \langle I_{L-1} \rangle_{\infty} & = -\Gamma (f + \langle \sigma^z_L \rangle_{\infty})    \label{currentboundaryright} \; .
\end{align}
This current is uniform across the chain, with its value dictated by the average magnetization at the boundaries; from now on, we remove the subscript and denote it simply by $\langle I \rangle_{\infty}$. In the absence of any bias ($f=0$), the boundary reservoirs induce an infinite temperature (maximally-mixed) steady state $\rho(\infty) = \mathbb{I}/2^L$, in which the values of magnetization and current all vanish.

\subsection{Steady-state ansatz}
\label{sec:steady_state_ansatz}
In order to study the transport properties of the model at the steady state, both numerically and analytically, we make use of a perturbative ansatz characterized by an expansion in terms of potential bias $f$ as \cite{Znidaric2010,Znidaric2011,Znidaric2013}
\begin{align}
    \label{ansatz}
    \rho & = \frac{1}{2^L} \big[\mathbb{I} + f (H + B) + \mathcal{O}(f^2) \big] \; , \\
    H & = \sum_{r=1}^{L}\sum_{j=1}^{L+1-r} h^{(r)}_j H^{(r)}_j \; , \\ 
    B & = \sum_{r=2}^{L}\sum_{j=1}^{L+1-r} b^{(r)}_j B^{(r)}_j \; ,
\end{align}
where $h_j^{(r)}$ and $b_j^{(r)}$ are expansion coefficients, and the corresponding operators are given by
\begin{align}
    \label{ansatzoperators}
    H^{(1)}_j &= - \sigma^{z}_j \hspace{2.5mm}, \nonumber \\
    H^{(r)}_j &= \sigma^{x}_j Z^{(r-2)}_{j+1} \sigma^{x}_{j+r-1} + \sigma^{y}_j Z^{(r-2)}_{j+1} \sigma^{y}_{j+r-1} \hspace{2.5mm}  , \hspace{2.5mm} r \geq 2 \nonumber \\
    B^{(r)}_j & = \sigma^{x}_j Z^{(r-2)}_{j+1} \sigma^{y}_{j+r-1} - \sigma^{y}_j Z^{(r-2)}_{j+1} \sigma^{x}_{j+r-1} \hspace{2.5mm}  , \hspace{2.5mm} r \geq 2 \; .
\end{align}
These operators are thus strings of Pauli matrices starting at site $j$ and having length $r$, where $Z^{(r-2)}_{k+1} = \sigma^{z}_{k+1} \sigma^{z}_{k+2}~...~ \sigma^{z}_{k+r-3} \sigma^{z}_{k+r-2}$ are strings of Pauli $z$ matrices of length $r-2$. The defining characteristic of this ansatz is that all operators in the expansion are orthogonal according to the Hilbert-Schmidt inner product $\langle A, B\rangle \equiv \mathrm{Tr}[A^{\dagger}B]$. This implies that the expectation values of $H^{(r)}_j$ and $B^{(r)}_j$ are \textit{exactly} determined within first order in $f$. In particular, $H_j^{(1)}$ and $B_j^{(2)}$ are respectively proportional to the magnetization and current
\begin{align}
    \label{observables-ansatz}
    \langle \sigma^{z}_{j} \rangle_{\infty} = -f h^{(1)}_j \hspace{2.5mm} , \hspace{2.5mm} \langle I_j \rangle_{\infty} = f~\frac{4J b^{(2)}_j}{\hbar} \; .
\end{align}
Once we determine the first order expansion coefficients $h^{(r)}_j$ and $b^{(r)}_j$, we have access to the exact current and magnetization profile \cite{Znidaric2010,Znidaric2011,Znidaric2013,Arenas2024}. In order to obtain them, we insert Eq.~\eqref{ansatz} into Eq.~\eqref{masterequation} and set the left-hand side to zero as demanded at stationarity (see Refs.~\cite{Znidaric2010,Znidaric2013} and Sec.~\ref{app:ansatz} for more details). This gives rise to a set of equations for the coefficients
\begin{align}
    & \hbar\Gamma(1+h^{(1)}_1) - 4J b_1^{(2)} = 0 \label{lineqs1}\; , \\
    & \hbar \Gamma(1-h^{(1)}_L) - 4J b_{L-1}^{(2)} = 0 \label{lineqs2} \; , \\
    & b_j^{(2)} - b_{j-1}^{(2)} = 0 \; , \hspace{2.5mm} j = 2,...,L-1 \label{lineqs3} \; ,
\end{align}
for $r = 1$. Note that Eqs.~\eqref{lineqs1}--\eqref{lineqs3} just express the fact that the current is uniform in the steady state, with its value dictated by the boundary reservoirs as discussed before. For $r \geq 2$ we have
\begin{align}
    \label{lineqs4}
    & J \big[ h_j^{(r-1)} - h_{j+1}^{(r-1)} + h_j^{(r+1)} - h_{j-1}^{(r+1)} \big] +  (\epsilon_{j} - \epsilon_{j+r-1})h_{j}^{(r)} + \hbar\Big[ \gamma + \frac{\Gamma}{4}(\delta_{j,1} + \delta_{j+r-1,L}) \Big] b^{(r)}_j = 0 \; , \\
    \label{lineqs5} 
    & J \big[b_j^{(r-1)} - b_{j+1}^{(r-1)}+ b_j^{(r+1)} - b_{j-1}^{(r+1)} \big] + (\epsilon_{j} - \epsilon_{j+r-1})b_{j}^{(r)} - \hbar \Big[ \gamma + \frac{\Gamma}{4}(\delta_{j,1} + \delta_{j+r-1,L}) \Big] h^{(r)}_j = 0 \; .
\end{align}
In total, Eqs.~\eqref{lineqs1}--\eqref{lineqs5} are a closed set of $L^2$ coupled equations and will be the starting point of our study. This set of equations is equivalent to a Lyapunov-type equation for the correlation matrix in the presence of dephasing \cite{Znidaric2013}. Instead of the exponential scaling with the system size, the quadratic scaling makes the problem computationally tractable. Moreover, as we show below, it also allow us to find analytical expressions for the current under some mild assumptions.

\section{Dephasing-assisted transport}
\label{sec:dat}

We now discuss separately the effect of the tilt and dephasing on the transport properties of the model, while their interplay is addressed in the next section. The bias $f$ of each reservoir at the boundaries induces transport along the chain. Intuitively, a similar effect is expected in the presence of an electric field which increases/decreases at each site at a rate $U$, or tilt between neighbour sites. If the tilt decreases from left to right, the local potentials are
\begin{align}
    \label{localpotential}
    \epsilon_j = - Uj \; .
\end{align}
Setting the bias to $f=1$ drives excitations in the same direction (see Fig.~\ref{fig:schematic}). Note that changing the sign of $f$ inverts the sign of the current and magnetization, while changing the sign of $U$ affects neither \cite{Arenas2024}. 

Since the tilt $U$ represents the energy difference between nearest neighbours in the chain, the total tilt $V=U(L-1)$ represents the difference in energy across the whole chain. This quantity diverges in the thermodynamic limit, so that a physically meaningful analysis requires fixing the total tilt $V$ and allowing $U$ to decrease with the system size $L$. As discussed in Secs.~\ref{sec:tridiagonal} and \ref{sec:interplay}, this procedure introduces a non-trivial dependence of the current with system size in the presence of dephasing: while for fixed $U$ the current is expected to decrease with the system size $L$, this will no longer be the case for fixed $V$.

\begin{figure*}[h!]
\centering
 \includegraphics[width=8.65cm]{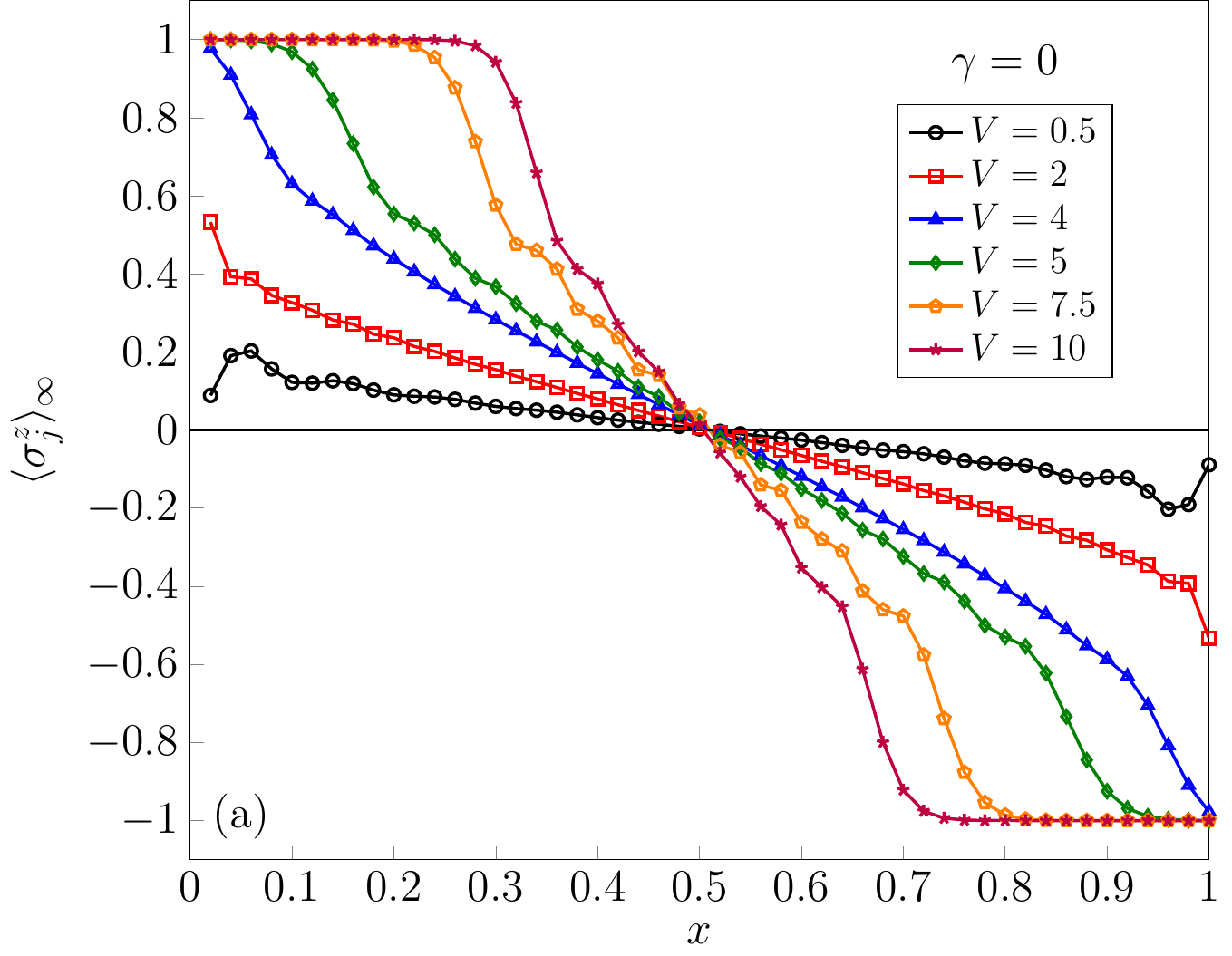}
 \includegraphics[width=8.8cm]{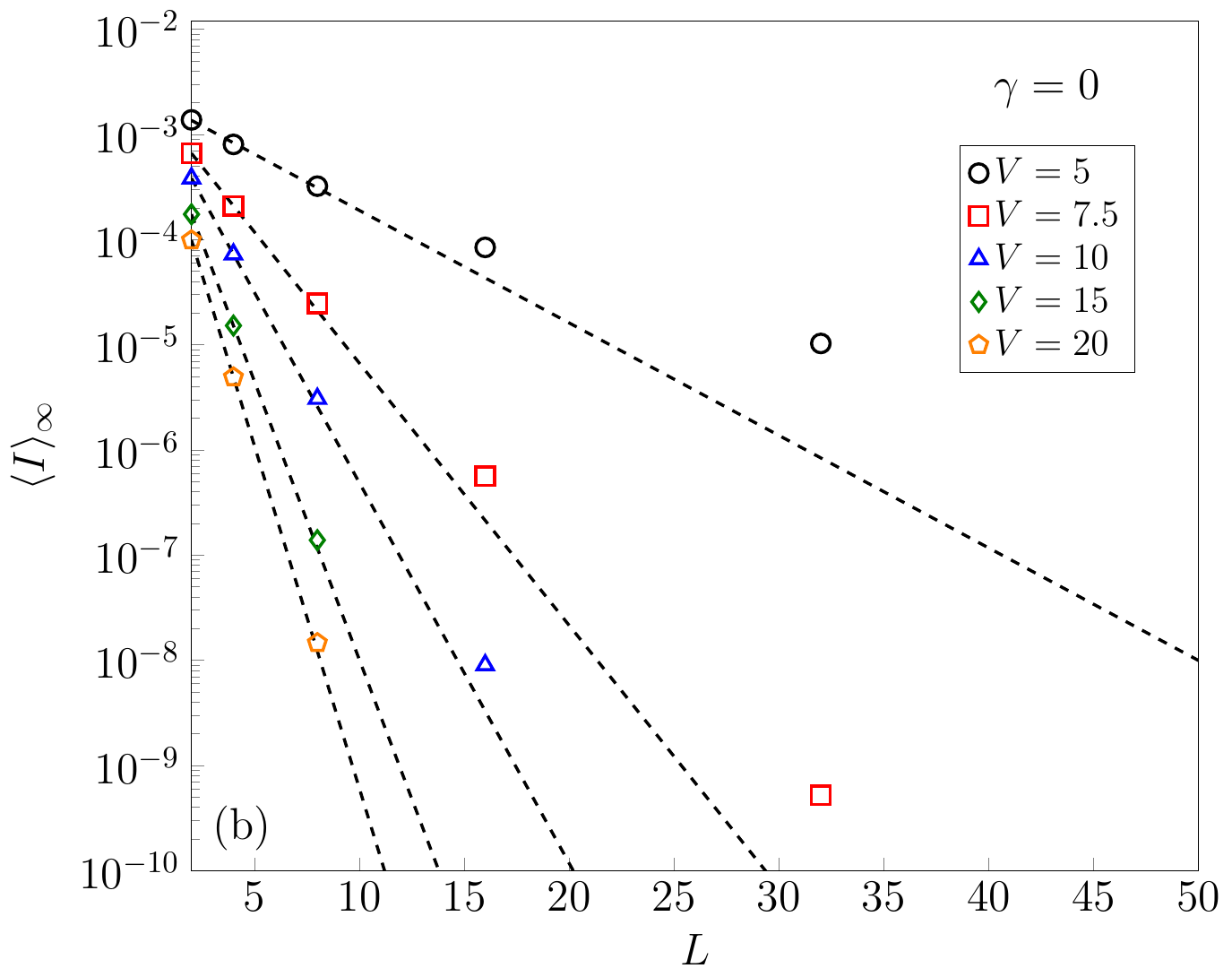}
\caption{(a) Steady-state magnetization profile $\{ \langle \sigma^z_j \rangle_{\infty} \}_{j=1}^{L}$ for system size $L=50$ and different values of total tilt $V$. The site positions are normalized $x=j/L$. (b) Steady-state current $\langle I \rangle_{\infty}$ as a function of system size for different values of total tilt $V$. The dashed lines represent exponential fits $\langle I \rangle_{\infty} \propto e^{-L/L_0}$ to the data with $L_0 = 4.06, 1.75, 1.20, 0.82, 0.67$ respectively for $V=5,7.5,10,15,20$. Parameters: $\hbar = f = J = 1$, $\Gamma = 0.01$ and $\gamma = 0$.}
\label{fig:localizedsystem}
\end{figure*}

Such a linear potential is known to induce Wannier-Stark localization in the absence of dephasing, with $U/\hbar$ being the frequency of Bloch oscillations \cite{Zener1934,Wannier1962,Hartmann2004,Guo2021}.
In the absence of dephasing $\gamma = 0$ and for non-vanishing tilt $V>0$, the steady state magnetization profile shows the formation of domain walls \cite{Arenas2024} -- regions of constant magnetization $\langle \sigma^z_j \rangle_{\infty} \simeq \pm f$ starting at the boundaries and extending inwards as shown in Fig.~\ref{fig:localizedsystem}(a). According to Eqs.~\eqref{currentboundaryleft} or \eqref{currentboundaryright}, this implies a vanishing current, which characterizes the localized system. Indeed we observe in Fig.~\ref{fig:localizedsystem}(b) that, as $V$ increases, the current displays an exponential decay with the system size $\langle I \rangle_{\infty} \sim e^{-L/L_0}$ with $L_0$ being the localization length. Note that $L_0$ here is different from the localization region of a single particle in a Wannier-Stark localized system given by $J/U$ \cite{Hartmann2004}. %We did not find any direct relationship between the two.

\begin{figure}[h!]
\centering
 \includegraphics[width=8.65cm]{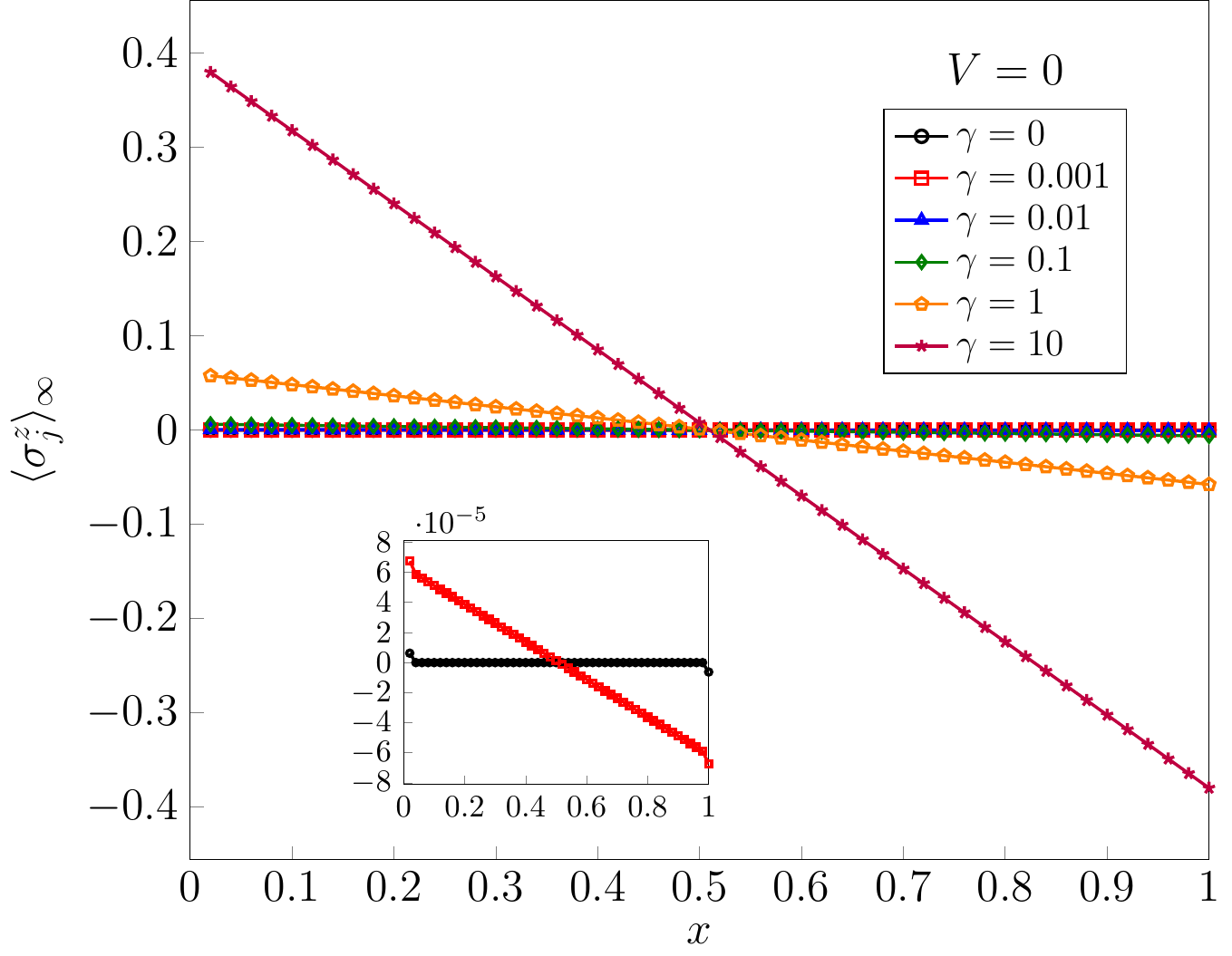}
\caption{Steady-state magnetization profile $\{ \langle \sigma^z_j \rangle_{\infty} \}_{j=1}^{L}$ for system size $L=50$ and different values of dephasing $\gamma$. The site positions are normalized $x=j/L$. The inset shows a zoom-in of the results for $\gamma=0$ and $\gamma=0.001$. Parameters: $\hbar = f = J = 1$, $V = 0$.}
\label{fig:notiltsystem}
\end{figure}

The results in the absence of tilt $U = V = 0$ are shown in Figs.~\ref{fig:notiltsystem} and \ref{fig:tridiagonal}(a), respectively for the magnetization and the current. We distinguish two regimes: $\gamma = 0$ and $\gamma > 0$. The first case corresponds to ballistic transport, characterized by a small (but non-zero) magnetization at the boundaries and zero bulk magnetization (inset of Fig.~\ref{fig:notiltsystem}), together with a size-independent current (see later Fig.~\ref{fig:tridiagonal}(a)). The second case corresponds to diffusive transport, characterized by an emerging linear magnetization profile as the dephasing rate increases, accompanied by a decreasing current in accordance with Eqs.~\eqref{currentboundaryleft} or \eqref{currentboundaryright}. For very large dephasing or system sizes the current scales as $\langle I \rangle_{\infty} \sim 1/(\gamma L) $ signaling diffusive transport. The exact expression for the current in the absence of tilt is given by Eq.~\eqref{currenttridiagonal} (with $U=V=0$) as we discuss in Sec.~\ref{sec:tridiagonal}.

\subsection{Exact current for small system sizes}

\begin{figure}[h!]
\centering
 \includegraphics[width=8.65cm]{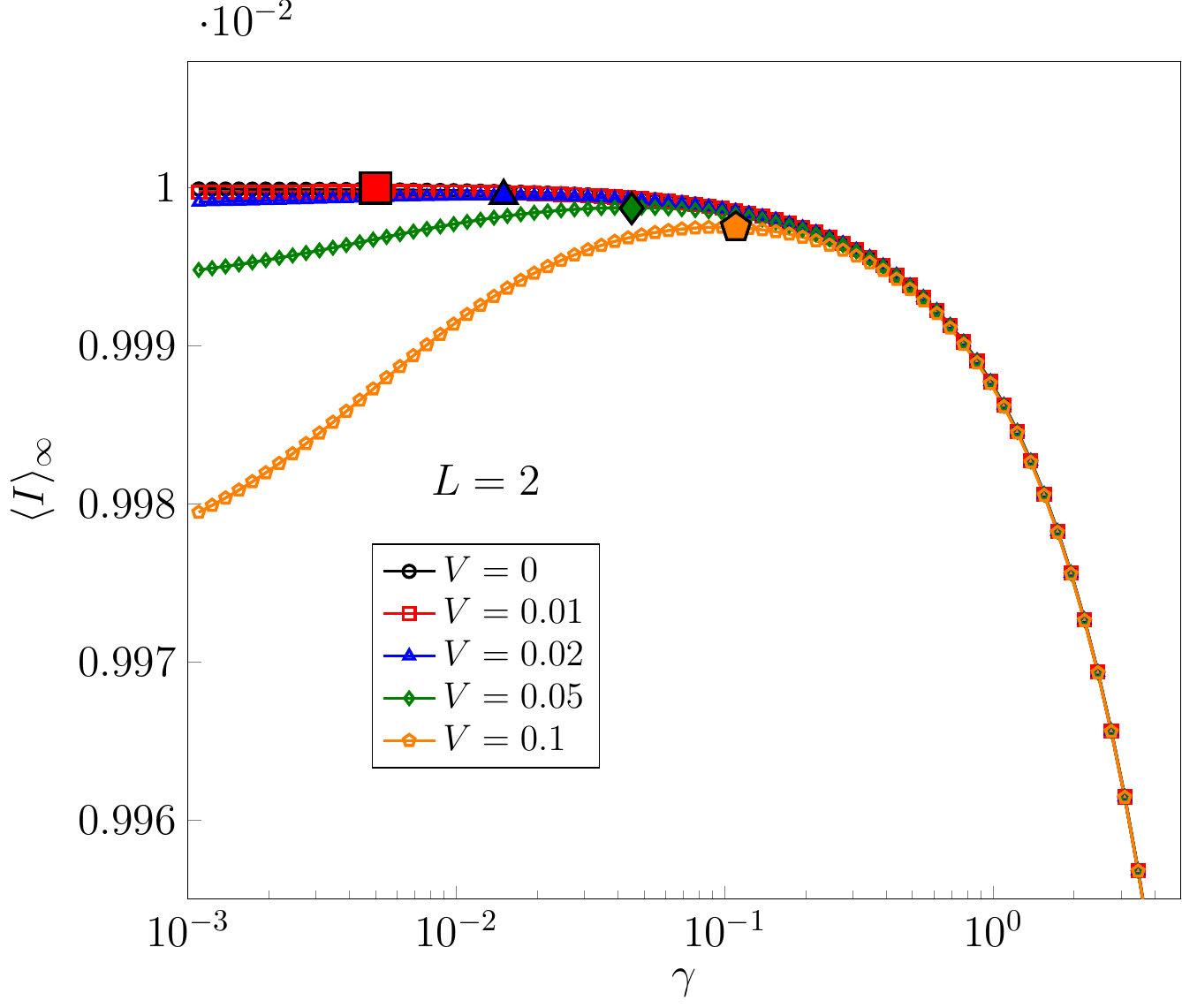}
\caption{Steady-state current $\langle I \rangle_{\infty}$ for $L=2$ as a function of dephasing $\gamma$ for different values of tilt $U=V$, computed from Eq.~\eqref{currentL2}. The large filled markers represent the maxima computed from Eq.~\eqref{dephmaxL2}. Parameters: $\hbar = f = J = 1$, $\Gamma = 0.01$.}
\label{fig:currentsmall}
\end{figure}

Having shown the degrading effect of the tilt and dephasing separately on the spin transport across the system, we now study their interplay which gives rise to dephasing-enhanced transport. In this section, we focus on transport for small system sizes $L=2,3$. An exact and compact solution for the current can be found from Eqs.~\eqref{lineqs1}--\eqref{lineqs5} in these cases. For $L = 2$ we obtain
\begin{align}
    \label{currentL2}
    \langle I \rangle_{\infty} = f~\frac{16 J^2 \Gamma (2 \gamma + \Gamma)}{16 J^2 (2 \gamma + \Gamma) + 4 \Gamma U^2 + \hbar^2\Gamma(2 \gamma + \Gamma)^2} \; .
\end{align}
Dephasing-assisted transport can be characterised through the maximum of Eq. \eqref{currentL2} with respect to the dephasing rate $\gamma$, yielding an optimal dephasing rate
\begin{align}
    \label{dephmaxL2}
    \gamma_{\mathrm{max}} = \frac{U}{\hbar} - \frac{\Gamma}{2} \geq 0 \; .
\end{align}
The behaviour of Eq.~\eqref{currentL2} is shown in Fig.~\ref{fig:currentsmall}. Note that since the dephasing is always positive, the maximum only occurs if $U/\hbar \geq \Gamma/2$.
For $L=3$ we have a similar expression, namely
\begin{align}
    \label{currentL3}
    \langle I \rangle_{\infty} = f~\frac{16 J^2 \Gamma (4 \gamma + \Gamma)}{16 J^2 (4 \gamma + \Gamma) + 16 \Gamma U^2 + \hbar^2\Gamma(4 \gamma + \Gamma)^2 } \; ,
\end{align}
where the maximum occurs at
\begin{align}
    \gamma_{\mathrm{max}} = \frac{U}{\hbar} - \frac{\Gamma}{4} \geq 0 \; .
\end{align}
The complexity of these expressions increases rapidly for $L > 3$, so the exact evaluation of the current and its maximum is too involved to be displayed here for larger systems. In the following section, we resort to an approximate solution for the current.

\subsection{Tridiagonal approximation \label{sec:tridiagonal}}

\begin{figure*}[h!]
\centering
    \includegraphics[width=8.6cm]{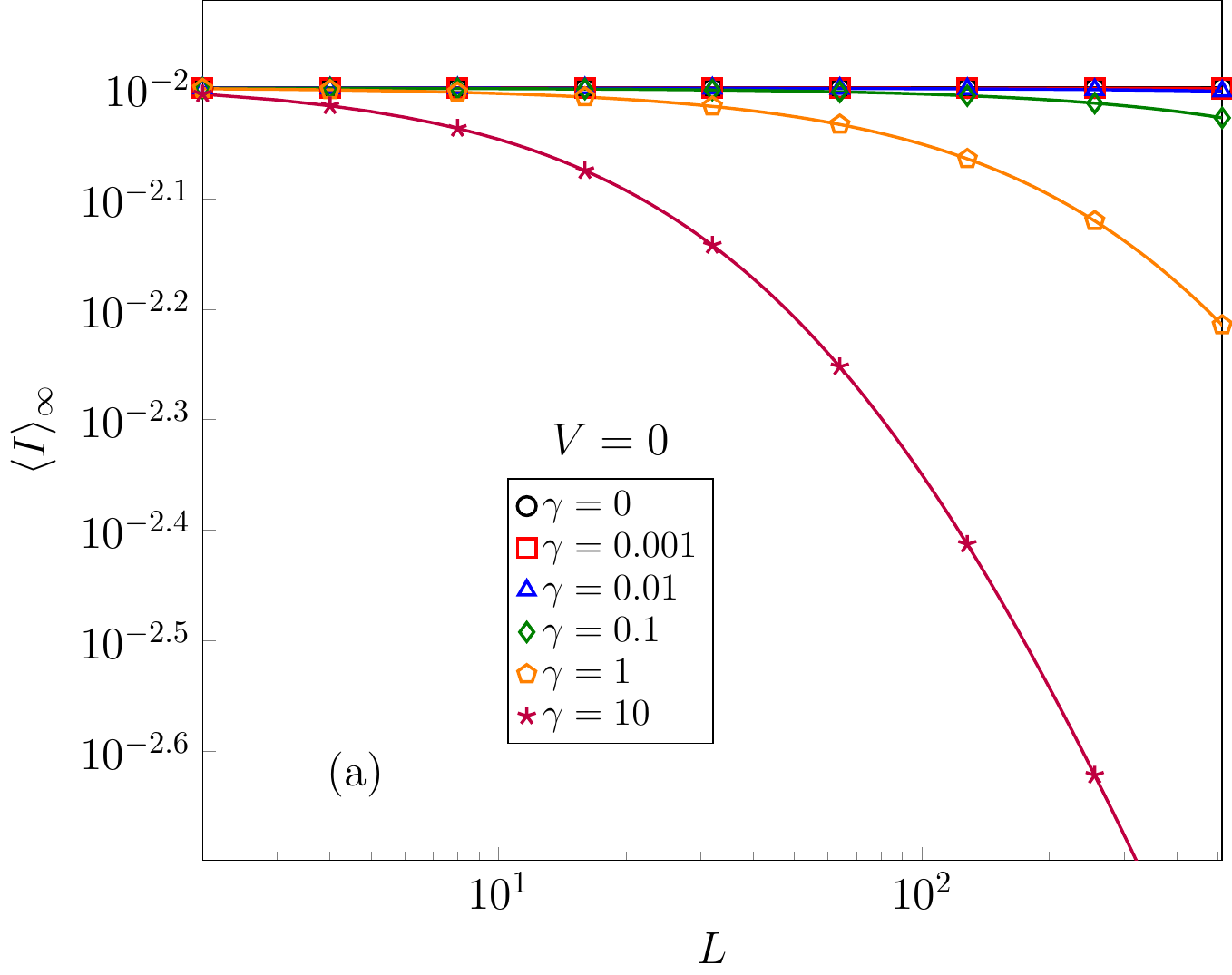}
    \includegraphics[width=8.6cm]{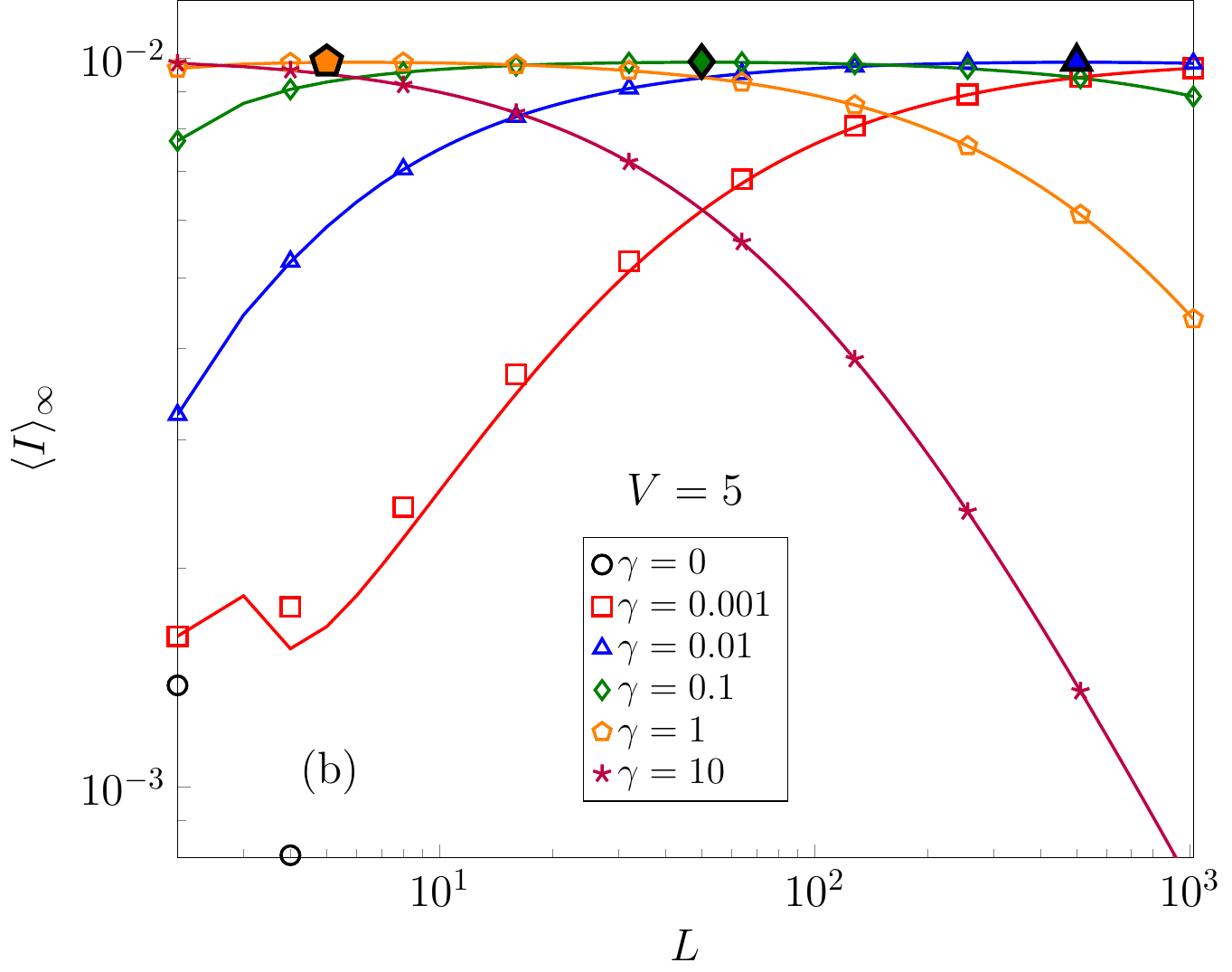}
    \includegraphics[width=8.6cm]{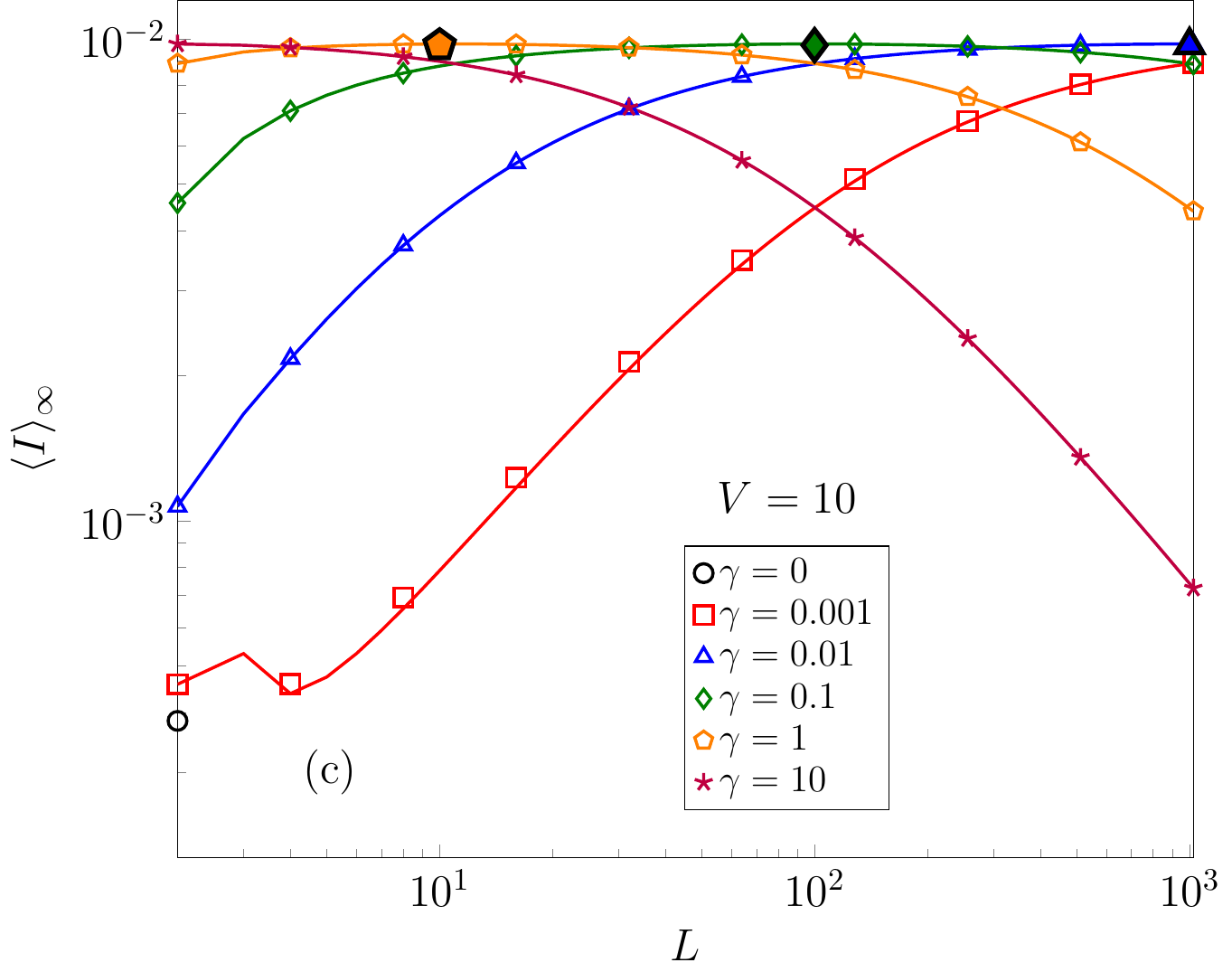}
    \includegraphics[width=8.6cm]{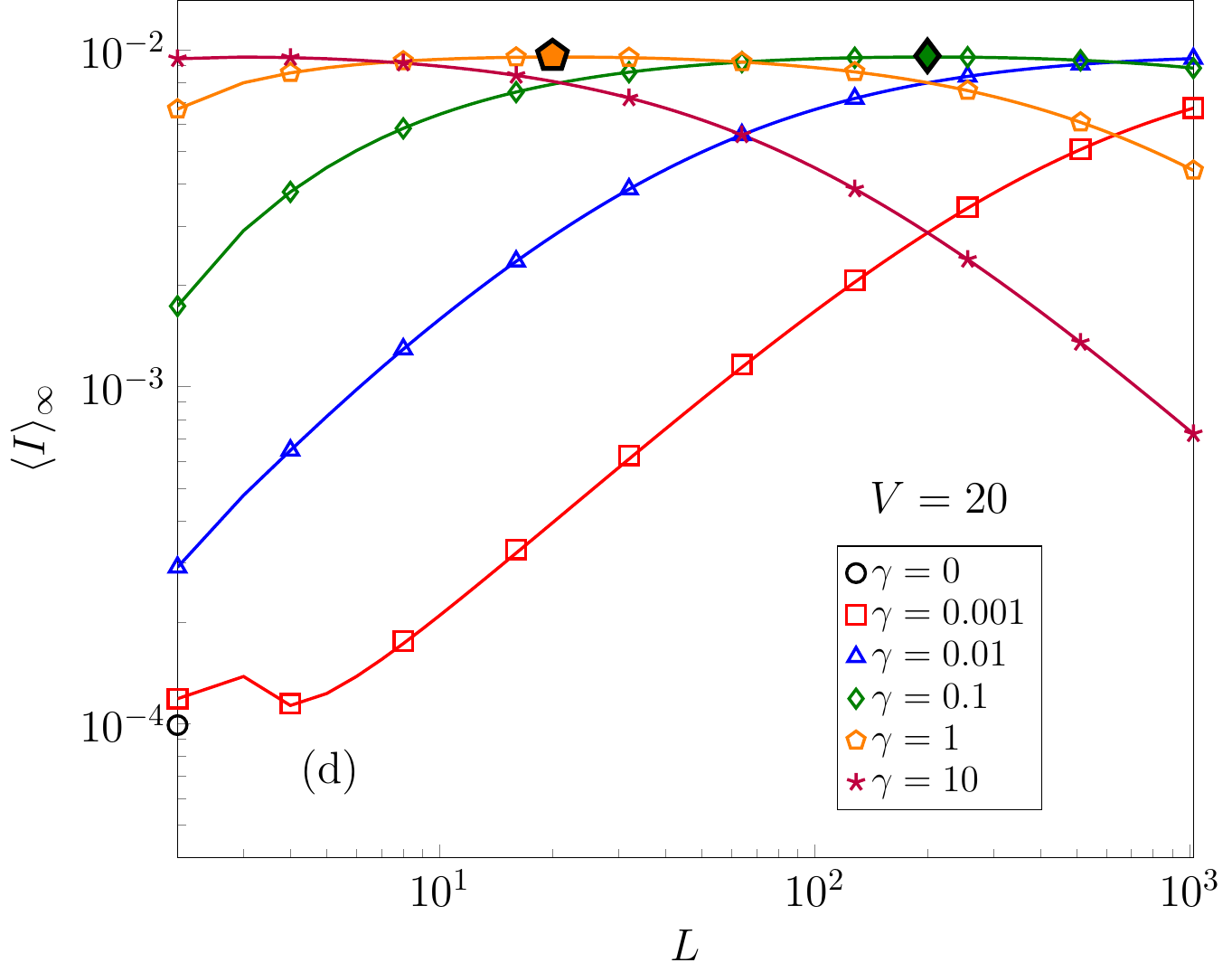}
\caption{Steady-state current $\langle I \rangle_{\infty}$ as a function of system size $L$ for different values of dephasing rate $\gamma$ and total tilt $V = U(L-1)$. (a) $V=0$, (b) $V=5$, (c) $V=10$ and (d) $V=20$. The unfilled markers represent the exact values computed from Eqs.~\eqref{lineqs1}--\eqref{lineqs5}; the lines represent the tridiagonal approximation in Eq.~\eqref{currenttridiagonal}. The large, filled markers represent the maximum of the current as a function of the system size occurring at $L_{\mathrm{max}} \simeq V/(\hbar \gamma)$. Parameters: $\hbar = f = J = 1$, $\Gamma = 0.01$.}
\label{fig:tridiagonal}
\end{figure*}

In order to find an exact solution for systems of arbitrary size, we perform the \textit{tridiagonal approximation} which consists of retaining strings of Pauli matrices up to $r=2$, that is, ignoring higher order coefficients in Eqs.~\eqref{lineqs1}--\eqref{lineqs5} so that $h^{(r)}_j,b^{(r)}_j = 0$ for $r \geq 3$. In the correlation matrix approach, this is equivalent to retaining only the diagonal and first off-diagonal elements \cite{Znidaric2013}. Physically, this implies that spin and current correlations can be ignored within first order in $f$, which allows the derivation of an exact expression for the current as a function of all parameters as we now show. We only have to consider Eqs.~\eqref{lineqs4} and \eqref{lineqs5} for $r=2$. Summing Eqs.~\eqref{lineqs4} from $j=1$ to $j=L-1$ we obtain
\begin{align}
    \label{der} 
    J(h^{(1)}_1 - h^{(1)}_{L}) & + \sum_{j=1}^{L-1} (\epsilon_j - \epsilon_{j+1}) h^{(2)}_{j} +  \hbar b [ \gamma(L-1) + \Gamma/4] = 0 \; ,
\end{align}
where $b \equiv b^{(2)}_j$ is the coefficient associated with the current, which is independent of $j$ as a consequence of Eq.~\eqref{lineqs3}. First, we can write $h^{(1)}_1 - h^{(1)}_{L}$ as a function of $b$ by using Eqs.~\eqref{lineqs1} and \eqref{lineqs2}, which yields
\begin{align}
    h^{(1)}_1 - h^{(1)}_{L} = \frac{8 J b}{\hbar \Gamma} - 2 \; .
\end{align}
Second, we can also express $h^{(2)}_{j}$ as a function of $b$ by using Eq.~\eqref{lineqs5}, which leads to
\begin{align}
    h^{(2)}_{j} = \frac{ (\epsilon_j - \epsilon_{j+1}) b}{\hbar[\gamma + (\delta_{j,1} + \delta_{j,L-1})\Gamma/4]} \; .
\end{align}
We can now substitute the previous two equations into Eq.~\eqref{der} and obtain an expression for $b$, namely
\begin{align}
    b = \frac{4J}{2 \hbar \gamma(L-1) + \hbar \Gamma + \frac{16J^2}{\hbar \Gamma}  + \frac{2}{\hbar} \sum_{j=1}^{L-1} \frac{(\epsilon_j - \epsilon_{j+1})^2}{\gamma +(\delta_{j,1} + \delta_{j,L-1})\Gamma/4}} \; .
\end{align}
The current is now obtained from Eq.~\eqref{observables-ansatz}. For the local tilted potential in Eq.~\eqref{localpotential} it becomes
\begin{align}
    \label{currenttridiagonal}
    \langle I \rangle_{\infty} = f~\frac{16J^2 \Gamma}{16J^2 + \hbar^2 \Gamma[\Gamma + 2 \gamma(L-1)] + 4 \Gamma U^2 
    \begin{cases}
      1/(2\gamma + \Gamma) & L = 2 \\
      (L-3)/(2\gamma) + 4/(4\gamma + \Gamma) & L \geq 3
    \end{cases} 
    } \; .
\end{align}
We note that this expression is exact for $L=2$ and $L=3$, in which case we recover Eqs.~\eqref{currentL2} and \eqref{currentL3}, respectively. It is also exact in the absence of tilt $U=V=0$ \cite{Znidaric2010,Znidaric2013,Turkeshi2021}, which is verified in Fig.~\ref{fig:tridiagonal}(a). In last three panels of Fig.~\ref{fig:tridiagonal} we plot the exact solution and the tridiagonal approximation for non-zero values of total tilt $V > 0$. We observe that the approximation works remarkably well, deviating from the exact solution only for very small dephasing rates and system sizes; when $\gamma = 0$, the approximation fails dramatically and only the exact solution is shown. The latter case corresponds to a Wannier-Stark localized system which we already discussed. Moreover, we see that for fixed $\gamma$ the current can now increase with the system size up to a critical size $L_{\mathrm{max}}$ where it is maximal, and only then decreases. In the same way, we see that for fixed $L$ there is a critical dephasing rate $\gamma_{\mathrm{max}}$ at which the current is maximal. 

\subsection{Interplay between dephasing and tilt \label{sec:interplay}}

\begin{figure*}[h!]
%\centering
 \includegraphics[width=8.7cm]{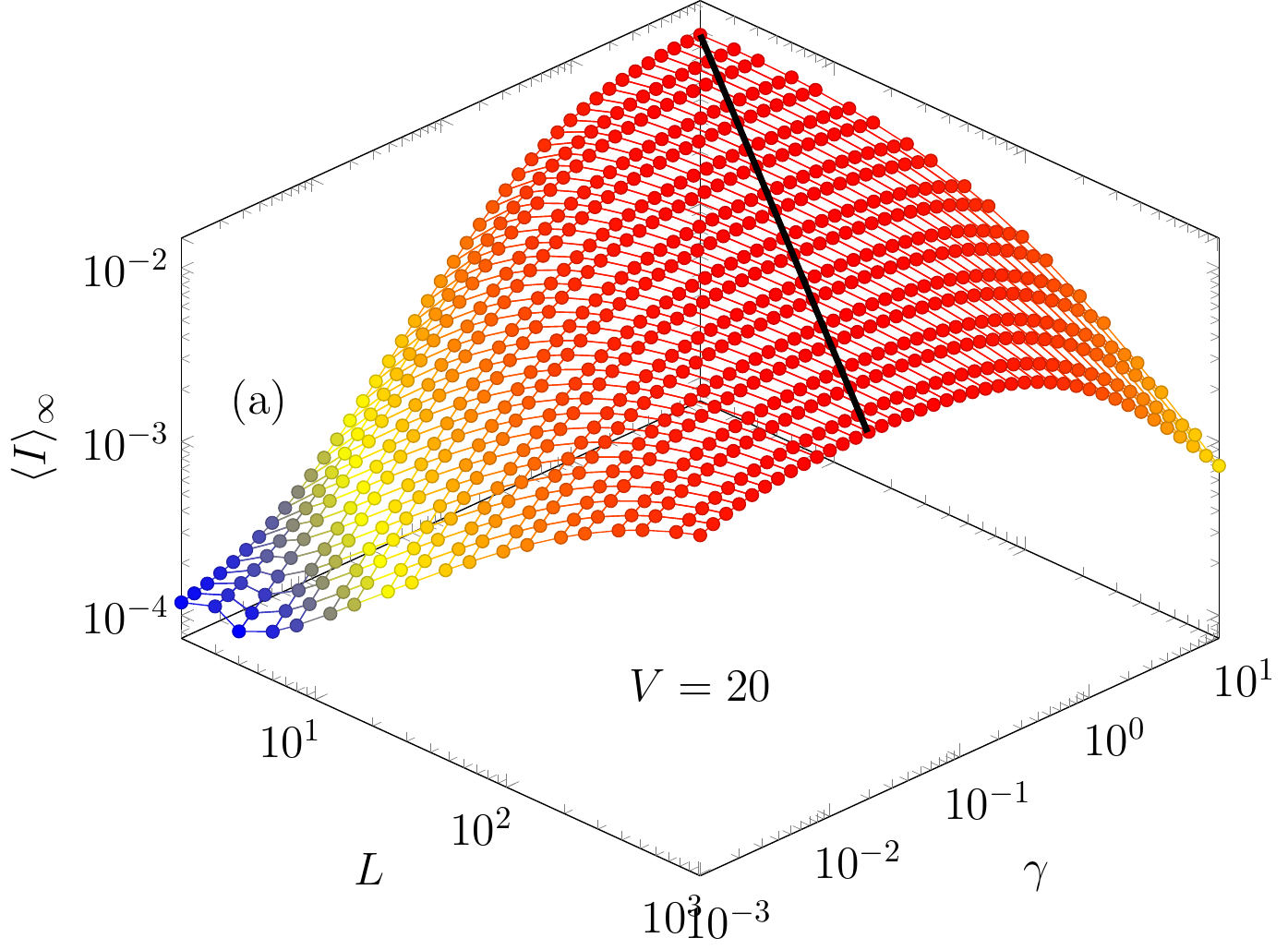}
    \includegraphics[width=8cm]{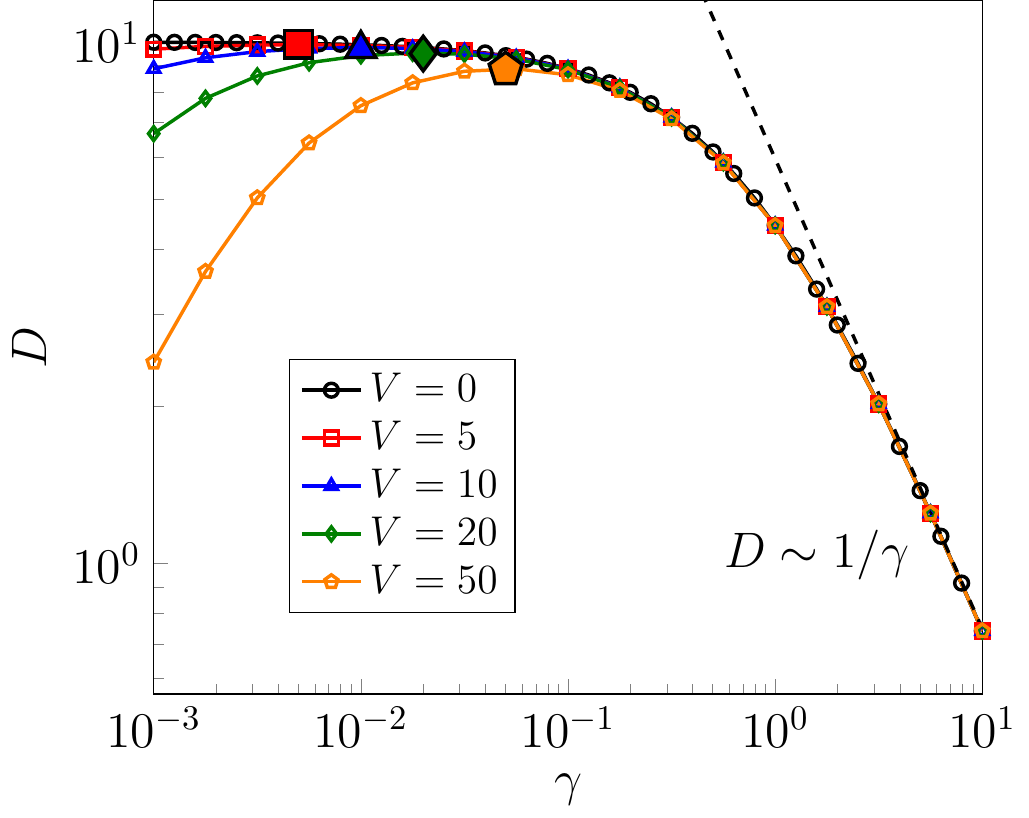}
\caption{(a) Exact steady-state current $\langle I \rangle_{\infty}$ as a function of system size $L$ and $\gamma$ for total tilt $V = 20$. The line represents the maximum current, obtained for dephasing values dictated by Eq.~\eqref{dephmax}. The colour scheme is a visual aid representing the value of the current increasing from blue to red. (b) Diffusion coefficient as a function of dephasing rate $\gamma$ for different values of total tilt $V$. The large, filled markers represent the maximum occurring at $\gamma_{\mathrm{max}} \simeq V/(\hbar L)$. The remaining parameters are the same as in previous figures.}
\label{fig:current3d}
\end{figure*}

Since the validity of Eq.~\eqref{currenttridiagonal} is established, we now use it to determine both $L_{\mathrm{max}}$ and $\gamma_{\mathrm{max}}$. The maxima of Eq.~\eqref{currenttridiagonal} as a function of $L$ (for fixed $V$ and $\gamma$) and $\gamma$ (for fixed $L$ and $V$) can be determined exactly. However, as before, the expressions are too lengthy to be displayed. More physically intuitive expressions are obtained by assuming that $L \gg 1$ and $L \gg 8 \gamma / (4\gamma + \Gamma)$ in the denominator of Eq.~\eqref{currenttridiagonal} which leads to the approximate maxima
\begin{align}
    L_{\mathrm{max}} & \simeq \frac{V}{\hbar \gamma} \label{Lmax} \; , \\
    \gamma_{\mathrm{max}} & \simeq \frac{V}{\hbar L} = \frac{U}{\hbar} \label{dephmax} \; .
\end{align}
These maxima characterize dephasing-assisted transport as we show in Fig.~\ref{fig:current3d}(a). 
Below the critical length ($L_{\mathrm{max}} \gg L \gg 1$), dephasing is not yet effective and an increase in $L$ decreases the local tilt $U = V/(L-1)$ and thus localization, which increases the current according to Eq.~\eqref{currenttridiagonal}. Above this length scale ($L \gg L_{\mathrm{max}} \gg 1$) dephasing becomes effective and diffusive transport clearly emerges $\langle I \rangle_{\infty} \sim 1/(\gamma L) $. Analogously, below the critical dephasing rate ($\gamma_{\mathrm{max}} \gg \gamma$) dephasing jump events happen at a much lower frequency than Bloch oscillations, so the current flowing remains hampered by localization. Increasing the rate of dephasing destroys localization and increases the current. When the dephasing rate passes this threshold ($\gamma \gg \gamma_{\mathrm{max}}$)  diffusive behaviour emerges. We illustrate this in Fig.~\ref{fig:current3d}(b) where we plot the (finite size) diffusion coefficient defined by $\langle I \rangle_{\infty} \equiv f D/L$ as a function of dephasing rate for the largest system size we reached $L = 1000$.

\begin{figure}[h!]
\centering
 \includegraphics[width=8.6cm]{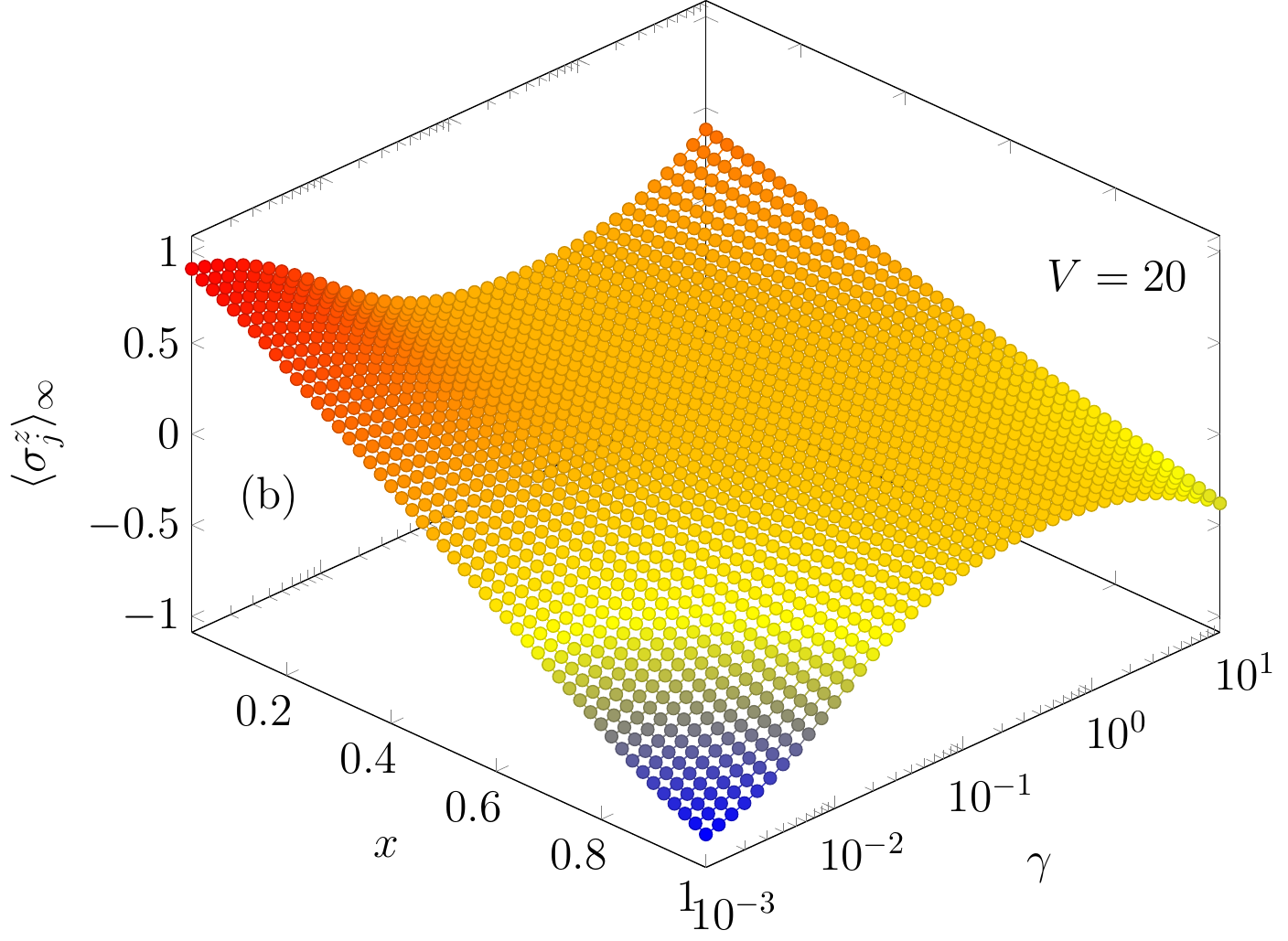}
\caption{Steady-state magnetization profile $\{ \langle \sigma^z_j \rangle_{\infty} \}_{j=1}^{L}$ as a function of dephasing $\gamma$ for system size $L=50$ and total tilt $V=20$. The site positions are normalized $x=j/L$. The colour scheme is a visual aid representing the value of the magnetization increasing from blue to red. The remaining parameters are the same as in previous figures.}
\label{fig:magnetization3d}
\end{figure}

The imprint of dephasing-enhanced transport is also evident in the magnetization profile, as shown in Fig.~\ref{fig:magnetization3d}. Here, varying the dephasing strength towards its critical value \eqref{dephmax} decreases the magnetization gradient, rendering the magnetization profile almost uniform. This is in contrast with the results obtained in the absence of tilt, where an increasing dephasing rate tends to increase the magnetization gradient and decrease uniformity as shown in Fig.~\ref{fig:notiltsystem}(a). Uniformity as a feature of environment-assisted transport has been pointed out for quantum systems subject to on-site disorder \cite{Harush2020}; our results confirm that the same phenomenon appears in the presence of a linear potential inducing Wannier-Stark localization.

\section{Conclusion and Outlook}
\label{sec:conclusion}
In this paper, we have exploited an exact ansatz to obtain the steady state of a non-interacting spin chain subject to bulk dephasing, a linear potential and boundary-driving. This description has allowed us to unravel the delicate interplay between Wannier-Stark localization and dephasing-induced diffusive transport. We applied the tridiagonal approximation, where only on-site and nearest-neighbour correlation matrix elements are retained, to derive an approximate expression for the steady-state current as a function of both dephasing and tilt. This expression is found to closely match the exact solution for a wide range of parameters, and reveals that the maximum current occurs for a dephasing rate equal to the period of Bloch oscillations in the Wannier-Stark localized system. Fixing the total tilt across the system then revealed a maximum in the current as a function of system size. This evidences a critical system size beyond which Bloch oscillations are suppressed by dephasing.

Our work motivates future analytical and numerical analysis on the interplay between dephasing and a tilted potential in more complex and richer systems, e.g. those including interactions between neighboring sites \cite{nosotros2013prb,nosotros2013jstat,Mendoza:2014}, time-periodic driving \cite{de2022prb,de2024arxiv}, or more realistic (non-Markovian) boundary-driving schemes \cite{Prior2010,Tamascelli2018,Strathearn2018nat,nosotros2020prx,archak2021prb,segal2023prxq}. Another promising research avenue is to study the sensing capacity of our quantum many-body system. Indeed, it has been recently shown that the non-equilibrium dynamics of Bloch oscillations can enhance the sensing capacity of an isolated quantum many-body system \cite{Manshouri2024}; it would be interesting to examine whether such an advantage is found for a quantum many-body system subject to dephasing, particularly in the dephasing-assisted transport regime. Finally, our results can be readily verified in current photonic \cite{ma2019nat}, ion-trap \cite{morong2021nat} and cold-atom \cite{krinner2017jpcm,scherg2021nat,amico2022arxiv} experimental platforms.

\section*{Funding}

The authors declare financial support was received for the research, authorship, and/or publication of this article. S.L.J. acknowledges the financial support from a Marie Skłodowska-Curie Fellowship (Grant No. 101103884). S.R.C. gratefully acknowledge financial support from UK's Engineering and Physical Sciences Research Council (EPSRC) under grant EP/T028424/1 which is part of an EPSRC-SFI joint project QuamNESS funded also by the SFI under the Frontier For the Future Program. J.G. is supported by a SFI Royal Society University Research Fellowship.

\section*{Acknowledgements}

The authors want to thank Gabriel T. Landi for useful discussions.

\section{Supplementary Material}
\label{app:suppmat}

\subsection{Heisenberg equation}
\label{app:heisenberg}

For completeness, we show here how to obtain Eq.~\eqref{heisenbergequation}. Differentiating $\langle \sigma^{z}_j \rangle(t) \equiv \mathrm{Tr}[\sigma_j \rho(t)]$ with respect to time and using Eq.~\eqref{masterequation} yields
\begin{align}
    \label{heisenbergapp}
    \frac{d \langle \sigma^z_j \rangle(t)}{dt} = \frac{i}{\hbar}~ \mathrm{Tr}\big[ [H,\sigma^z_j] \rho(t) \big] + \sum_{\alpha = l,r,d} \mathrm{Tr} \big[ \sigma^z_j \mathcal{L}^{\alpha}[\rho(t)] \big] \; . 
\end{align}
Throughout this section, we make use of several relations for the Pauli matrices, which can all be deduced from their commutation and anti-commutation relations $[\sigma^{\alpha},\sigma^{\beta}] = 2i \varepsilon_{\alpha \beta \gamma} \sigma^{\gamma}$ and $\{\sigma^{\alpha},\sigma^{\beta}\} = 2 \delta_{\alpha \beta} \mathbb{I}_2$, respectively. Here $\alpha, \beta, \gamma \in \{ x, y, z \}$, $\mathbb{I}_2$ is the identity operator for a two-level system and $\epsilon_{\alpha\beta\gamma}$ is the Levi-Civita symbol with the Einstein summation notation being implicit. The commutator in Eq.~\eqref{heisenbergapp} can then be easily computed
\begin{align}
    \label{currentoperatorderivation}
    \frac{i}{\hbar}[H,\sigma^{z}_j] & = \frac{iJ}{\hbar} \sum_{k=1}^{L-1}[ \sigma_{k}^x \sigma^{x}_{k+1} + \sigma_{k}^y \sigma^{y}_{k+1},\sigma^{z}_j] \nonumber \\ 
    & =  \frac{2J}{\hbar} (\sigma_{j-1}^{x} \sigma_{j}^{y} - \sigma_{j-1}^{y} \sigma_{j}^{x}) - \frac{2J}{\hbar} (\sigma_{j}^{x} \sigma_{j+1}^{y} - \sigma_{j}^{y} \sigma_{j+1}^{x}) \nonumber \\
    & = I_{j-1} - I_{j} \; ,
\end{align}
In the first line only the kinetic energy part of the Hamiltonian is non-vanishing. In the second line, only the pairs $(k,k+1)$ which overlap with $j$ survive. In the last line, we identified the current operator in Eq.~\eqref{currentoperator}. Regarding the last term in Eq.~\eqref{heisenbergapp}, we can make progress by considering the dual superoperator acting on $\sigma^z_j$ defined by $\mathrm{Tr} \big[ \sigma^z_j \mathcal{L}^{\alpha}[\rho(t)] \big] = \mathrm{Tr} \big[ \mathcal{L}^{\alpha \dagger}[\sigma^z_j] \rho(t) \big]$ with
\begin{align}
    \mathcal{L}^{\alpha \dagger}[\cdot] = \sum_{j,\beta} L^{\alpha \beta \dagger}_j \cdot L_j^{\alpha \beta} - \frac{1}{2} \big\{ L_j^{\alpha \beta \dagger} L^{\alpha \beta}_j, \cdot \big\} \; .
\end{align}
The dephasing superoperator is equal to its dual and reads explicitly
\begin{align}
    \label{deph_superoperatorD}
    \mathcal{L}^{d}[\cdot] = \frac{\gamma}{2} \sum_{j=1}^{L} ( \sigma_j^z \cdot \sigma_j^z -~\cdot ) = \mathcal{L}^{d \dagger}[\cdot] \; .
\end{align}
It is straightforward to see that its action vanishes when applied to any Pauli $z$-matrix. Therefore dephasing does not contribute to Eq.~\eqref{heisenbergapp}. For the reservoir at the left boundary, we have the dual superoperator
\begin{align}
    \label{dualsuperoperatorBL}
    \mathcal{L}^{l \dagger}[\cdot] & = \frac{\Gamma(1+f)}{2} O_1^{l+ \dagger}[\cdot]~+~\frac{\Gamma(1-f)}{2} O_1^{l- \dagger}[\cdot] \\
    O_1^{l\pm \dagger}[\cdot] & \equiv \sigma_1^{\mp} \cdot \sigma_1^{\pm} - \frac{1}{2} \{\sigma_1^{\mp} \sigma_1^{\pm},\cdot \} \; ,
\end{align}
and a similar expression holds for the right reservoir in Eq.~\eqref{boundaryoperators}. Applying these superoperators to the Pauli z-matrix yields after some manipulation
\begin{align}
    \mathcal{L}^{l \dagger}[\sigma^z_j] & = \delta_{j,1} \Gamma(f - \sigma^z_1) \; , \\
    \mathcal{L}^{r \dagger}[\sigma^z_j] & = - \delta_{j,L} \Gamma(f + \sigma^z_L) \; .
\end{align}
Using the last two expressions and Eq.~\eqref{currentoperatorderivation} in Eq.~\eqref{heisenbergapp} immediately leads to Eq.~\eqref{heisenbergequation}.

\subsection{Steady-state ansatz}
\label{app:ansatz}

In order to derive Eqs.~\eqref{lineqs1}--\eqref{lineqs5}, we insert Eq.~\eqref{ansatz} into Eq.~\eqref{masterequation} and set the left-hand side to zero as demanded at steady state. This gives
\begin{align}
    \label{ansatzapp}
    0 & = \frac{1}{2^L} \sum_{\alpha = l,r,d,0} \big[ \mathcal{L}^{\mathrm{\alpha}}[\mathbb{I}] + f ( \mathcal{L}^{\mathrm{\alpha}}[H] + \mathcal{L}^{\mathrm{\alpha}}[B])] \; , \\
    \mathcal{L}^{\mathrm{\alpha}}[H] & = \sum_{r=1}^{L}\sum_{j=1}^{L+1-r} h^{(r)}_j \mathcal{L}^{\mathrm{\alpha}}[H^{(r)}_j] \; , \\ 
    \mathcal{L}^{\mathrm{\alpha}}[B] & = \sum_{r=2}^{L}\sum_{j=1}^{L+1-r} b^{(r)}_j \mathcal{L}^{\mathrm{\alpha}}[B^{(r)}_j] \; .
\end{align}
In the last expression we ignore all terms beyond first order in $f$ and defined the superoperator for free evolution
\begin{align}
    \label{superoperatorH}
    \mathcal{L}^{0} [\cdot] \equiv -\frac{i}{\hbar}[H,~\cdot~] \; .
\end{align}

\subsubsection{Action of the dephasing reservoir}

We start by computing the action of the dephasing superoperator which reads explicitly
\begin{align}
    \label{superoperatorD}
    \mathcal{L}^{d}[\cdot] = \frac{\gamma}{2} \sum_{j=1}^{L} ( \sigma_j^z \cdot \sigma_j^z -~\cdot ) \equiv \frac{\gamma}{2} \sum_{j} O^{d}_j[\cdot] \; .
\end{align}
It is straightforward to see that the action of this superoperator vanishes when applied to $\mathbb{I}$ and $\sigma^{z}$. For $r \geq 2$, we have to compute terms of the following form
\begin{align}
    \mathcal{L}^{d}\Big[\sigma^{\alpha}_j Z^{(r-2)}_{j+1} \sigma^{\beta}_{j+r-1} \Big] & = \frac{\gamma}{2} \Big[O^{d}_j[\sigma^{\alpha}_j]~Z^{(r-2)}_{j+1} \sigma_{j+r-1}^{\beta} + \sigma^{\alpha}_j Z^{(r-2)}_{j+1} O^{d}_{j+r-1} [\sigma_{j+r-1}^{\beta}] \Big] \; ,
\end{align}
Note that the right-hand side has only two non-vanishing terms on the boundaries since the action on $\sigma^z$ vanishes. We then have
\begin{align}
    O^{d}_j[\sigma^\alpha_j] = \sigma_j^z \sigma_j^{\alpha} \sigma_j^z - \sigma_j^{\alpha} = - 2 \sigma_j^{\alpha}  (\delta_{\alpha,y} + \delta_{\alpha,x}) \; .
\end{align}
Using this last expression and carrying through simple algebraic manipulation, the action of the dephasing is written as
\begin{align}
    \mathcal{L}^{d}[\mathbb{I}] & = \mathcal{L}^{d}[H_j^{(1)}] = 0 \; , \\
    \mathcal{L}^{d}[H_j^{(r)}] & = -2 \gamma H_j^{(r)} \hspace{2.5mm} , \hspace{2.5mm} r \geq 2 \\
    \mathcal{L}^{d}[B_j^{(r)}] & = -2 \gamma B_j^{(r)} \hspace{2.5mm} , \hspace{2.5mm} r \geq 2 \; .
\end{align}

\subsubsection{Action of the boundary reservoirs}

We first illustrate the action for the left boundary reservoir. The superoperator is given by
\begin{align}
    \label{superoperatorBL}
    \mathcal{L}^{l}[\cdot] & = \frac{\Gamma(1+f)}{2} O_1^{l+}[\cdot]~+~\frac{\Gamma(1-f)}{2} O_1^{l-}[\cdot] \; , \\
    O_1^{l\pm}[\cdot] & \equiv \sigma_1^{\pm} \cdot \sigma_1^{\mp} - \frac{1}{2} \{\sigma_1^{\mp} \sigma_1^{\pm},\cdot \} \; .
\end{align}
In opposition to dephasing, this superoperator does not preserve the identity
\begin{align}
    O_1^{l\pm}[\mathbb{I}] = \pm \sigma_1^{z} = \mp H^{(1)}_1 \; .
\end{align}
As a consequence, its action on the Pauli matrices $\sigma_j^{z}$ depends on $j$. Namely, we obtain
\begin{align}
    O_1^{l\pm}[\sigma_j^{z}] = 
    \begin{cases}
      - \sigma^z_1 \; , & j = 1 \\
      \pm \sigma_1^{z} \sigma^z_j \; , & j = 2,3...L \; .
    \end{cases} 
\end{align}
Inserting these results into Eq.~\eqref{superoperatorBL} we obtain the actions
\begin{align}
    \mathcal{L}^{l}[\mathbb{I}] & = -\Gamma f H^{(1)}_1 \; , \\
    \mathcal{L}^{l}[H^{(1)}_j] & = 
    \begin{cases}
        - \Gamma H_1^{(1)} & j = 1 \\
        f \Gamma H_1^{(1)} H_j^{(1)} & j = 2,3,\dots,L \; .
    \end{cases}
\end{align}
Inspecting the last expression together with Eq.~\eqref{ansatzapp}, we see that for $j=2,3,..L$ the reservoir creates terms in the expansion which are proportional to $f^2$ instead of $f$ -- they represent higher order spin correlations which do not affect the steady state values of magnetization and current. We henceforth ignore these second order terms by considering the action of the reservoir on Pauli matrices located at the reservoir site. Proceeding with the calculations for $r \geq 2$, we have to compute terms of the form
\begin{align}
    \mathcal{L}^{l} \Big[\sigma^{\alpha}_j Z^{(r-2)}_{j+1} & \sigma^{\beta}_{j+r-1} \Big] = \delta_{j,1} \Bigg\{ \frac{\Gamma(1+f)}{2} O^{l+}_1[\sigma^{\alpha}_1] + \frac{\Gamma(1-f)}{2} O^{l-}_1[\sigma^{\alpha}_1] \Bigg\}~Z^{(r-2)}_{j+1} \sigma^{\beta}_{j+r-1} \; .
\end{align}
It is easy to verify that
\begin{align}
    O^{l \pm}_1[\sigma^{\alpha}_1] = -\frac{\sigma_1^{\alpha}}{2} \hspace{2.5mm} , \hspace{2.5mm} \alpha = x,y \; .       
\end{align}
Using the last two expressions and carrying out the algebra, the results for the left reservoir to first order in $f$ are
\begin{align}
    \mathcal{L}^{l}[\mathbb{I}] & = - \Gamma f H^{(1)}_1 \; , \\
    \mathcal{L}^{l}[H^{(1)}_j] & = -\delta_{j,1} \Gamma H^{(1)}_j \; , \\
    \mathcal{L}^{l}[H^{(r)}_j] & = - \delta_{j,1} \frac{\Gamma}{2} H_j^{(r)} \hspace{2.5mm} , \hspace{2.5mm} r \geq 2 \; , \\
    \mathcal{L}^{l}[B^{(r)}_j] & = - \delta_{j,1} \frac{\Gamma}{2} B_j^{(r)} \hspace{2.5mm} , \hspace{2.5mm} r \geq 2 \; .
\end{align}
Proceeding the same way for the right boundary reservoir we obtain the result
\begin{align}
    \mathcal{L}^{r}[\mathbb{I}] & = \Gamma f H^{(1)}_L \; ,\\
    \mathcal{L}^{r}[H^{(1)}_j] & = -\delta_{j,L} \Gamma H^{(1)}_j \; , \\
    \mathcal{L}^{r}[H^{(r)}_j] & = - \delta_{j+r-1,L} \frac{\Gamma}{2} H_j^{(r)} \hspace{2.5mm} , \hspace{2.5mm} r \geq 2 \; , \\
    \mathcal{L}^{r}[B^{(r)}_j] & = - \delta_{j+r-1,L} \frac{\Gamma}{2} B_j^{(r)} \hspace{2.5mm} , \hspace{2.5mm} r \geq 2 \; .
\end{align}

\subsubsection{Action of Hamiltonian}

Regarding the Hamiltonian superoperator in Eq.~\eqref{superoperatorH}, it is easy to see that it preserves the identity $\mathcal{L}^{0}[\mathbb{I}] = 0$ while $\mathcal{L}^{0}[H_j^{(1)}]$ yields the well-known result
\begin{align}
    \mathcal{L}^{0}[H_j^{(1)}] = - \frac{2J}{\hbar}(B^{(2)}_j - B^{(2)}_{j-1}) \; .
\end{align}
This result was already implicitly used in Eq.~\eqref{currentoperatorderivation} and simply expresses that the change in magnetization at a given site is due to differences between incoming and outgoing currents at that site. For $r \geq 2$, it is useful to write the superoperator  
as a sum of two contributions $\mathcal{L}^{0} \equiv \mathcal{L}^{0_1} + \mathcal{L}^{0_2}$, where
\begin{align}
    \mathcal{L}^{0_1} \cdot & \equiv -\frac{i J}{\hbar} \sum_{j=1}^{L-1}~\Big\{ [\sigma_j^x \sigma_{j+1}^x,\cdot] + [\sigma_j^y \sigma_{j+1}^y,\cdot] \Big\} \; , \\ \mathcal{L}^{0_2} \cdot & \equiv -\frac{i}{\hbar} \sum_{j=1}^{L}~\epsilon_j [\sigma_j^z,\cdot] \; .
\end{align}
We compute only the action of $\mathcal{L}^{0_1}$ and the one for $\mathcal{L}^{0_2}$ is obtained in a similar manner. We first evaluate operators for $r = 2$ which contain no Pauli $z$-strings (see Eq.~\eqref{ansatzoperators} for the exact expressions). It involves computing terms of the form
\begin{align}
    \sum_{j=1}^{L-1} [\sigma_j^{\gamma} \sigma_{j+1}^{\gamma}, 
 \sigma_k^{\alpha} \sigma_{k+1}^{\beta}] & = \sigma^{\gamma}_{k-1}[\sigma_k^{\gamma},\sigma_k^{\alpha}]\sigma_{k+1}^{\beta}+ [\sigma_k^{\gamma} \sigma_{k+1}^{\gamma},\sigma_k^{\alpha} \sigma_{k+1}^{\beta}] + \sigma^{\alpha}_{k}[\sigma_{k+1}^{\gamma},\sigma_{k+1}^{\beta}]\sigma_{k+2}^{\gamma} \nonumber \\
    & = 2i~[\varepsilon_{\gamma \alpha \omega} \sigma_{k-1}^{\gamma} \sigma_k^{\omega}\sigma_{k+1}^{\beta} + \varepsilon_{\alpha \beta \omega}(\delta_{\gamma \alpha} \sigma_{k+1}^{\omega} - \delta_{\gamma \beta} \sigma_{k}^{\omega}) + \varepsilon_{\gamma \beta \omega} \sigma_{k}^{\alpha} \sigma_{k+1}^{\omega}\sigma_{k+2}^{\gamma}] \; .
\end{align}
In the first line, the only non-vanishing terms are for those pairs $(j,j+1)$ that partially or completely overlap with $(k,k+1)$. In the second line, we used the standard properties of the Pauli matrices. The last expression is enough to compute the action of $\mathcal{L}^{0_1}$ on $H^{(2)}_k$ and $B^{(2)}_k$ which reads
\begin{align}
    \mathcal{L}^{0_1}[H^{(2)}_j] & = -\frac{2J}{\hbar}(B_j^{(3)} - B_{j-1}^{(3)}) \; , \\
     \mathcal{L}^{0_1}[B^{(2)}_j] & = -\frac{4J}{\hbar}(H_{j+1}^{(1)} - H_{j}^{(1)})-\frac{2J}{\hbar}(H_j^{(3)} - H_{j-1}^{(3)}) \; .
\end{align}
For $r \geq 3$, the operators in Eq.~\eqref{ansatzoperators} have Pauli-$z$ strings. We need to compute terms of the form
\begin{align}
    \sum_{j=1}^{L-1} &[\sigma_j^{\gamma} \sigma_{j+1}^{\gamma},\sigma_k^{\alpha} Z^{(r-2)}_{k+1}\sigma_{k+r-1}^{\beta}] = \sigma_{k-1}^{\gamma} [\sigma_k^{\gamma},\sigma_k^{\alpha}] Z^{(r-2)}_{k+1} \sigma_{k+r-1}^{\beta} + [\sigma_k^{\gamma} \sigma_{k+1}^{\gamma},\sigma_k^{\alpha} \sigma_{k+1}^{z}] Z^{(r-2)}_{k+2} \sigma_{k+r-1}^{\beta} \nonumber \\
    & + \sigma_k^{\alpha} Z^{(r-2)}_{k+1}[\sigma^{\gamma}_{k+r-1},\sigma^{\beta}_{k+r-1}]\sigma^{\gamma}_{k+r} + \sigma_k^{\alpha} Z^{(r-3)}_{k+1} [\sigma_{k+r-2}^{\gamma} \sigma_{k+r-1}^{\gamma},\sigma_{k+r-2}^{z} \sigma_{k+r-1}^{\beta}] \; , \nonumber \\
    & = 2i~\big[ \varepsilon_{\gamma \alpha \omega} \sigma_{k-1}^{\gamma} \sigma_k^{\omega} Z^{(r-2)}_{k+1} \sigma_{k+r-1}^{\beta} + \varepsilon_{\gamma \beta \omega} \sigma_k^{\alpha} Z^{(r-2)}_{k+1} \sigma_{k+r-1}^{\omega} \sigma_{k+r}^{\gamma} \nonumber \\
    & + (\varepsilon_{\gamma \alpha \omega} \delta_{\gamma z} \sigma_k^{\omega} + \varepsilon_{\gamma z \omega} \delta_{\gamma \alpha} \sigma_{k+1}^{\omega}) Z^{(r-3)}_{k+2} \sigma_{k+r-1}^{\beta} + \sigma_{k}^{\alpha} Z^{(r-3)}_{k+1} (\varepsilon_{\gamma z \omega} \delta_{\gamma \beta} \sigma_{k+r-2}^{\omega} + \varepsilon_{\gamma \beta z} \delta_{\gamma z} \sigma_{k+r-1}^{\omega})\big] \; .
\end{align}
Again, the only non-vanishing terms are for those pairs $(j,j+1)$ that partially or completely overlap with the string $(k,k+1,...k+r-1)$. No commutator appears containing pairs of $\sigma_z$ since these also vanish. The last expression suffices to compute the action of $\mathcal{L}^{0_1}$ on $H^{(r)}_k$ and $B^{(r)}_k$ for $r \geq 3$ and a similar reasoning holds for $\mathcal{L}^{0_2}$. We summarize the results below:
\begin{align}
    \mathcal{L}^{0}[\mathbb{I}] & = 0 \; , \\
    \mathcal{L}^{0}[H_j^{(1)}] & = - \frac{2J}{\hbar}(B^{(2)}_j - B^{(2)}_{j-1}) \; , \\
    \mathcal{L}^{0}[B_j^{(2)}] & = \frac{4J}{\hbar}(H_{j}^{(1)} - H_{j+1}^{(1)})+\frac{2J}{\hbar}(H_j^{(3)} - H_{j-1}^{(3)}) + \frac{2(\epsilon_j - \epsilon_{j+1})}{\hbar}H_{j}^{(2)} \; , \\
    \mathcal{L}^{0}[H_j^{(r)}] & = - \frac{2J}{\hbar}\big[ (B_j^{(r-1)} - B_{j+1}^{(r-1)}) + (B_j^{(r+1)} - B_{j-1}^{(r+1)}) \big] - \frac{2(\epsilon_j - \epsilon_{j+r-1})}{\hbar}B_{j}^{(r)} \hspace{2.5mm} , \hspace{2.5mm} r \geq 3 \; , \\
    \mathcal{L}^{0}[B_j^{(r)}] & = \frac{2J}{\hbar}\big[ (H_j^{(r-1)} - H_{j+1}^{(r-1)}) + (H_j^{(r+1)} - H_{j-1}^{(r+1)}) \big] + \frac{2(\epsilon_j - \epsilon_{j+r-1})}{\hbar}H_{j}^{(r)} \hspace{2.5mm} , \hspace{2.5mm} r \geq 3 \; .
\end{align}
With the action of all superoperators computed and using Eq.~\eqref{ansatzapp} we obtain the set of Eqs.~\eqref{lineqs1}--\eqref{lineqs5} for the coefficients used in the main text.

\bibliographystyle{Frontiers-Harvard}
\bibliography{references}

\end{document}